\documentclass[%
 reprint,
superscriptaddress,
 amsmath,amssymb,
 aps,
]{revtex4-1}

\usepackage{graphicx}
\usepackage{dcolumn}
\usepackage{bm}
\usepackage{xcolor}

\begin{document}

\preprint{APS}

\title{High-Q  transparency band in all-dielectric metasurfaces induced by a
quasi bound state in the continuum 
}

\author{Diego R. Abujetas}

 \email{diego.romero@iem.cfmac.csic.es}
\affiliation{%
 Instituto de Estructura de la Materia (IEM-CSIC), Consejo Superior de Investigaciones Cient\'{\i}ficas,\\ Serrano 121, 28006 Madrid, Spain
}%
\author{\'Angela Barreda}
\affiliation{%
 Institute of Applied Physics, Abbe Center of Photonics, Friedrich Schiller University of Jena, Albert-Einstein-Str. 15, 07745 Jena, Germany.
}%
\author{Fernando Moreno}%
\affiliation{%
 Department of Applied Physics, University of Cantabria, Santander, Cantabria, 39005, Spain.
}%
\author{Amelie Litman}
\author{Jean-Michel Geffrin}%
\affiliation{%
 Aix Marseille Université, CNRS, Centrale Marseille, Institut Fresnel, Marseille, France
}%
\author{Jos\'e A. S\'anchez-Gil}%
 \email{j.sanchez@csic.es}
\affiliation{%
 Instituto de Estructura de la Materia (IEM-CSIC), Consejo Superior de Investigaciones Cient\'{\i}ficas,\\ Serrano 121, 28006 Madrid, Spain
}%

\date{\today}

\begin{abstract}
Bound states in the continuum (BICs) emerge throughout Physics as leaky/resonant modes that remain however highly localized. They have attracted much attention in Optics and Photonics, and especially in metasurfaces, i.e. planar arrays  of sub-wavelength meta-atoms. One of their most outstanding feature is the arbitrarily large Q-factors they induce upon approaching the BIC condition, which we exploit here to achieve a narrow transparency band. We first show how to shift a canonical BIC in an all-dielectric metasurface, consisting of high-refractive disks exhibiting in- and  out-of-plane magnetic dipole (MD) resonances, by tuning the periodicity of the array. By means of our coupled electric/magnetic dipole formulation, we show analytically that when the quasi-BIC overlaps with the broad  (in-plane MD)  resonance,  a full transparency band emerges with diverging Q-factor upon approaching the BIC condition in parameter space.  
Finally,  our experimental measurements in the microwave regime with a large array of  high-refractive-index disks confirm the theoretical predictions. Our results reveal a simple mechanism to engineer an ultra-narrow BIC-induced transparency band that could be exploited throughout the electromagnetic spectrum with obvious applications in filtering and sensing.
\end{abstract}

\maketitle


\section{\label{sec:Intro}Introduction}

Bound states in the continuum (BICs), ubiquitous in Physics as a wave phenomena, are attracting much attention in recent years \cite{Hsu2016a}. These are states that, despite lying in the continuum of modes, remain localized. Such properties make them particularly interesting in Photonics \cite{Marinica2008,Hsu2013,Bulgakov2014,Koshelev2018,Koshelev2019a}, for their diverging Q-factors without the need of complex fine optical cavities, thus leading to a rich phenomenology being explored these days such as lasing \cite{Kodigala2017,Ha2018}, enhanced non-linearities \cite{Carletti2018} and photoluminescence \cite{Zhu2020}, sensing \cite{Yesilkoy2019}, and spin-directive coupling  \cite{Zito2019}. Interestingly, upon slightly perturbing the parameter that governs the BIC condition, high-Q resonances are observed, termed in turn quasi-BICs \cite{Taghizadeh2017,Timofeev2018,Abujetas2019c}, which, if interacting with another broad resonance,  may as well induce Fano resonances with extremely narrow, asymmetric line shapes \cite{Rybin2017,Abujetas2017,Koshelev2018,Abujetas2019a,Bogdanov2019,Abujetas2019c}. 

Among the variety of the configurations where BICs emerge, much effort has focused on so called metasurfaces; namely, planar arrays with sub-wavelength periodicity such that only the zero-order specular reflection/transmission is allowed \cite{Holloway2012a,Glybovski2016,Li2017c,Neshev2018,Qiao2018,Sun2019,Kupriianov2019,Paniagua-Dominguez2019,Staude2019}. By limiting the outgoing radiation channels only to the specular one, various mechanisms may preclude radiation of localized/leaky modes through this channel, such as symmetry protection or accidental parameter tuning \cite{Hsu2016a,Koshelev2018,Koshelev2019a}. One of the mechanisms leading to symmetry-protected BICs at the $\Gamma$-point stems from the fact that vertical (out-of-plane) dipolar resonances are forbidden at normal incidence, termed also Brewster-like BICs. This has been shown with high-refractive-index (HRI), all-dielectric meta-atoms (disks o pillars) exhibiting either electric dipole resonances in the visible \cite{Ha2018} or magnetic dipole resonances in the GHz \cite{Abujetas2019d}. The latter indeed showed how a canonical isolated BIC emerges from a single, non-degenerate magnetic-dipole resonance, thanks to the fact that this is the lowest-order (non-overlapping) resonance of the high-refractive-index disk meta-atoms used therein.
On the other hand, recall the classical analogue of quantum electromagnetically-induced transparency (EIT) has been widely explored in metasurfaces, where the three level system needed is typically replaced by mode coupling of two resonances of various kinds  \cite{Zhang2008,Kurter2011,Gu2012,Tassin2012a,Fedotov2013,Zhu2013,Yang2014c,Schaafsma2016,Manjappa2016,Halpin2017a,Yahiaoui2018a,Shamkhi2019a}. Nonetheless, the use of BICs to achieve ultra-narrow EIT has not been explored thus far. 

In this regard, we will show below how this canonical BIC can be shifted by tuning the periodicity of the array, while keeping its "metasurface" character, so as to make it overlap  with a broad resonance stemming from the second lowest-order (in-plane magnetic dipole) resonance, leading to a quasi BIC-induced transparency (BIT) band with diverging Q-factor. In Sec~\ref{sec:MD}, we exploit our coupled electric/magnetic dipole formulation~\cite{Abujetas2020x} to explore the lattice-induced shift of both magnetic dipole resonances to achieve such overlap and to demonstrate analytically that the asymmetry factor vanishes exactly when they fully overlap (meaning full transparency). Section~\ref{sec:FANOBIC} demonstrates the emergence of  BIT through analytical and numerical calculations of the reflectance from the high-refractive-index disk metasurface with lattice-period tuned accordingly. Our predictions will be validated in Sec.~\ref{sec:EXP} by microwave experimental measurements with a large, but finite, array of such high-refractive-index disks. Finally, our concluding remarks are included in Sec.~\ref{sec:Conc}.

\section{\label{sec:MD}Lattice-induced tuning of MD BIC}
\begin{figure}
\includegraphics[width=0.95\columnwidth]{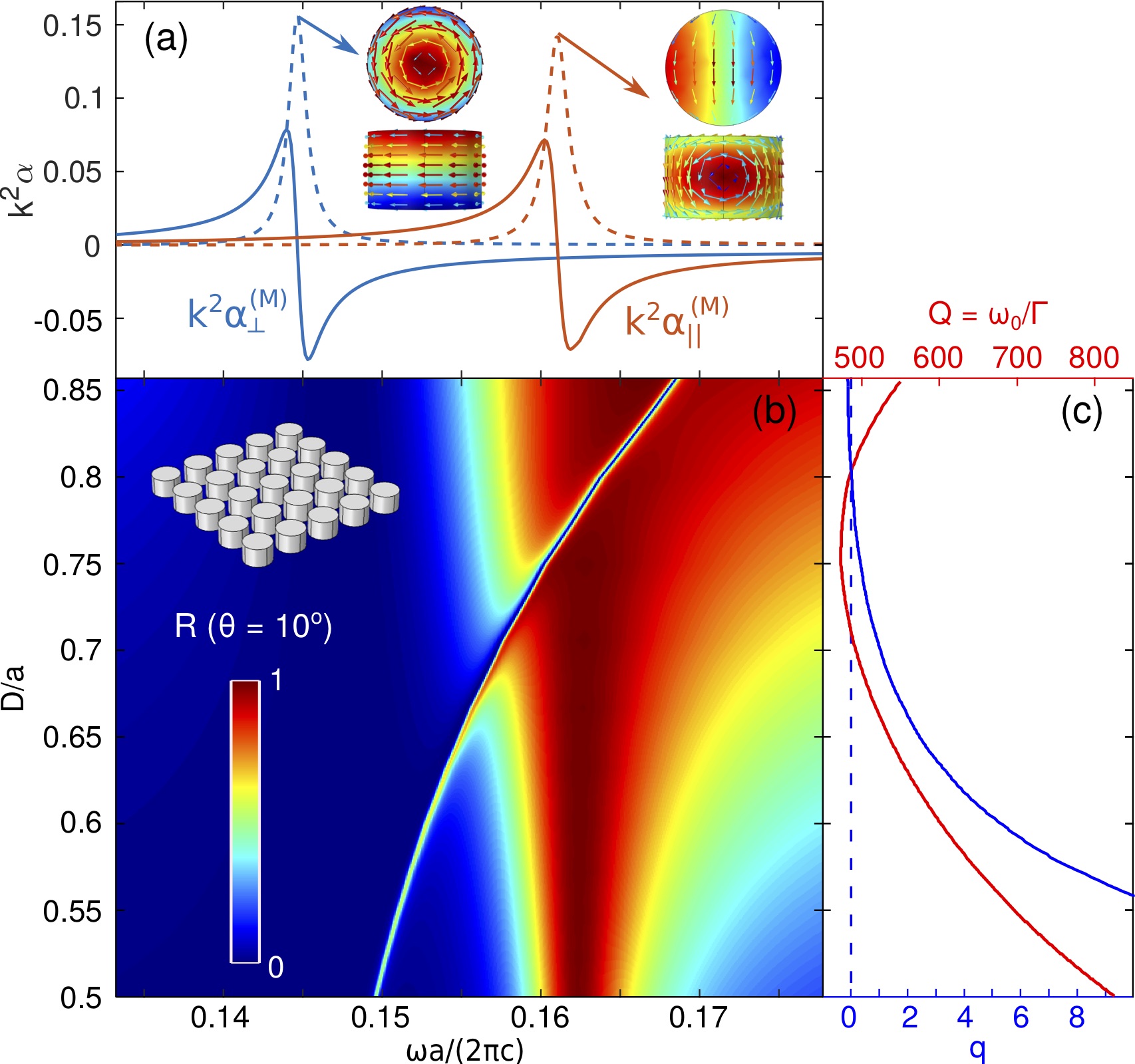}
\caption{(a)  Magnetic polarizibilities of a disk (aspect ratio=diameter/height$=D/L=1.5$) with dielectric constant of $\epsilon=78 + 0.05\mathrm i$ for different incident wave polarization and disk orientations extracted from SCUFF numerical results \protect{\cite{SCUFF1,SCUFF2}}. Insets: numerical calculations of the  magnetic charge (color maps) with electric currents (arrows) on the disk surface (top and side views) at the resonance frequencies. 
(b) CEMD calculations of the spectral dependence on the normalized reciprocal lattice vector $D/a$ of the in-  and out-of-plane MD resonances when the disks are arranged in an infinite square lattice with period $a$. Inset shows the disk metasurface geometry. (c) Asymmetry ($q$) and quality ($Q$) factors as a function of $D/a$ from a fit of the line shapes in (b) to the canonical Fano Eq. (1).
\label{fig:MD}}
\end{figure}
First of all, let us recall the magnetic-dipole (MD) resonances that the disks to be used in the experimental measurements exhibit \cite{Abujetas2019d}. If we assume a nearly dispertionless dielectric permittivity $\epsilon=78 + 0.05 \mathrm i$,
disks can be simply defined in terms of their aspect ratio $D/L=1.5$. The analysis can be thus generalized to most of the electromagnetic spectrum, since there are various materials that possess such dielectric constant at diverse spectral regions from the IR to the microwave regimes \cite{Paniagua-Dominguez2015}.
Bearing in mind that such disks will be placed in a planar array with their axis perpendicular to the array plane (see inset in Fig.~\ref{fig:MD}b), and we need in-plane and out-of-plane MD polarizibilities, we extract them from the scattering cross sections for the two relevant  angles of incidence in s polarization through SCUFF \cite{SCUFF1,SCUFF2} (free software implementation based on the method of moments), as shown in Fig.~\ref{fig:MD}a. 
The lowest-order one at $\omega a/c=1.45(2\pi)$ corresponds to the out-of-plane MD resonance, MD$_{\perp}$, which can be excited when the wave impinges off-axis with the magnetic field polarized along the cylinder axis; the second lowest-order one at $\omega a/c=1.61(2\pi)$ corresponds to the in-plane MD resonance, MD$_{||}$,  being excited at normal incidence with both polarizations.

We now explore the impact on both MD resonances of the periodicity $a$ when the disks are arranged in a  planar array (inset in Fig.~\ref{fig:MD}b), with the aim of finding an overlapping spectral regime, keeping within the non-diffracting regime where only the zero-order specular reflection/transmission appears. This is calculated through our coupled electric/magnetic dipole (CEMD) formulation for an infinite planar array as shown in \cite{Abujetas2020x,Abujetas2018a}; to this end, the MD polarizabilities shown in Fig.~\ref{fig:MD}a (and also the electric dipole, ED, ones) are needed.
The results are shown in Fig.~\ref{fig:MD}b as a function of the lattice reciprocal vector $G=2\pi/a$ normalized to the disk diameter as $GD/(2\pi)=D/a $. Note that, as the lattice period diminishes, so that disks become closer to each other, the MD$_{\perp}$ is blue-shifted from its position for an isolated disk, whereas the MD$_{||}$ varies more slowly with lattice periodicity. 

This is somewhat expected since the MD$_{\perp}$ resonance exhibits stronger radiation within the plane array, which corresponds to the equator of the MD  emission. Recall that it is the coupling of  such radiated fields in between disks which governs such blueshift. This can be verify through the electromagnetic charges and currents on the disk surface, numerically calculated in Fig.~\ref{fig:MD}a: in-plane electric fields circulating inside the disks, so that  magnetic charges accumulate on both planar faces of the disks (inducing a strong vertical MD), whereas the normal electric fields are stronger on the disk side. The MD$_{||}$ resonance, by contrast, presents stronger electric fields circulating over the disk rectangular cross sections, induced an in-plane MD perpendicular to the disk axis, thus showing weaker in-plane MD interaction between disks. 
\begin{figure}
\includegraphics[width=\columnwidth]{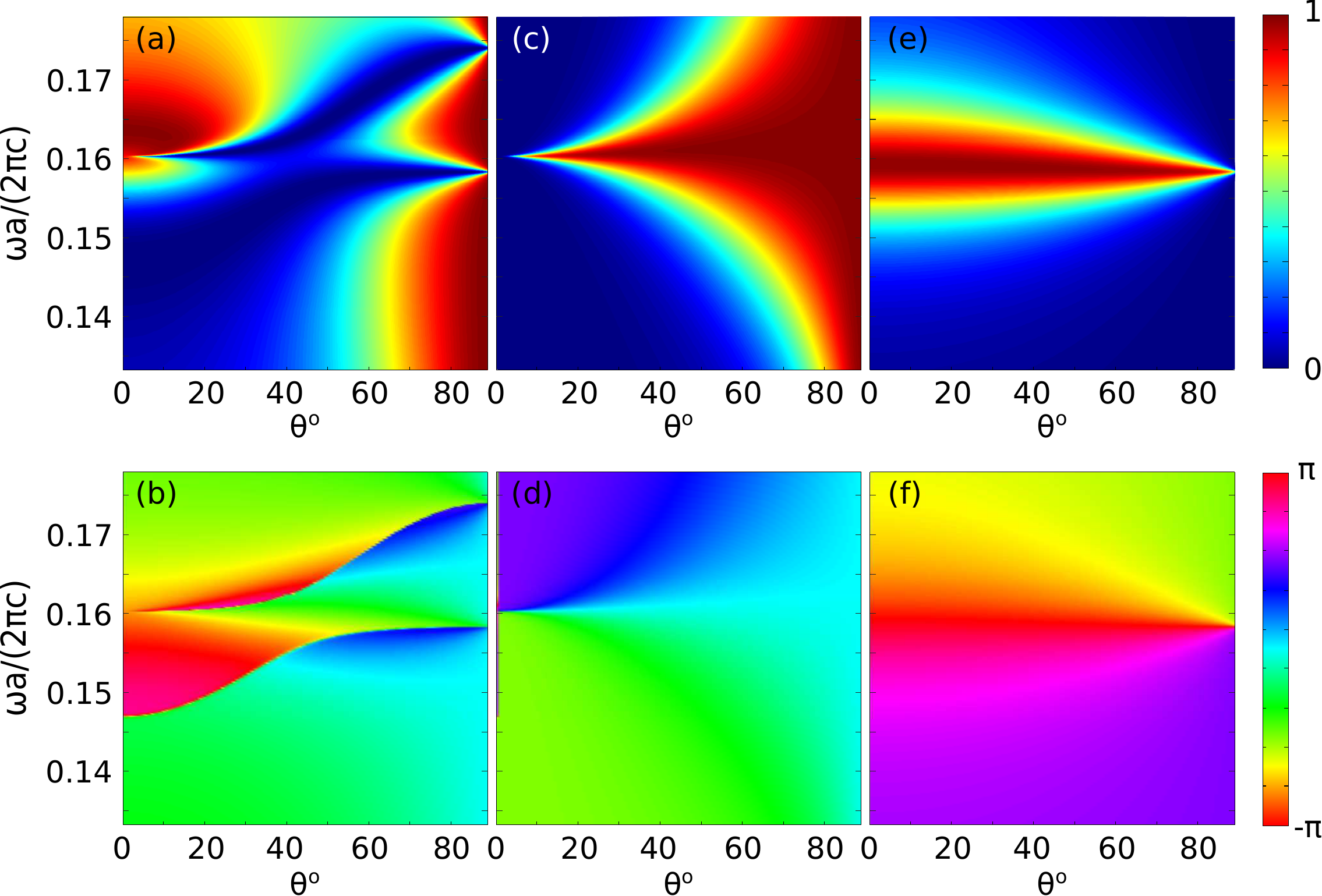}
\caption{Contour maps of the s-polarized reflectance $R(\omega,\theta$) intensity (a,c,e) and phase (b,d,f) from an infinite square array (lattice period $D/a=0.75$)
as a function of angle of incidence $\theta$ and normalized frequency $\omega a/c$, theoretically calculated through CEMD using the electric/magnetic polarizibilities  obtained from the SCUFF numerical calculations of the dielectric resonator disks in Fig. 1a, including separately the contributions from each MD polarizability: (c,d) out-of-plane MD$_{\perp}$ and (e,f) in-plane (MD$_{||}$).
\label{fig:CEMD}}
\end{figure}
We next investigate analytically the interference of both MD resonant bands. From our CEMD theory, reflectance can be written as a canonical Fano resonance:
\begin{equation}
R_0(\nu,\theta) \approx \sigma(\varepsilon)\frac{(\varepsilon+q)^2}{\varepsilon^2+1}
\label{Eq:Fano}\end{equation}
where $\varepsilon=(\omega-\omega_0)/(\Gamma/2)$ is the normalized frequency, $\omega_0$ being the BIC frequency and $\Gamma$ its FWHM. All magnitudes can be related to our CEMD formulation \cite{Abujetas2018a} through dressed polarizibilities $\tilde{\alpha}$ (given by bare polarizibilities and matrix elements of the lattice Green's function, $\tilde{\alpha}_{\beta}^{-1}=\alpha_{\beta}^{-1}-G_{b\beta\beta}, \beta=x,y,z$), as follows:
\begin{subequations}
\begin{eqnarray}
&& \sigma(\varepsilon)=(2k_za^2)^{-1}|\tilde{\alpha}_{||}\cos^2\theta|^2\\
&& \epsilon = \frac{\Re(1/\tilde{\alpha}_{\perp})}{\Im(1/\tilde{\alpha}_{\perp})} \\
&&q= - \frac{\Re(1/\tilde{\alpha}_{||})}{\Im(1/\tilde{\alpha}_{||})} 
\end{eqnarray}\label{Eq:FanoCEMD}
\end{subequations}
Let us now analyze them in detail. $\Gamma$ is nearly constant due to the fact that the angle of incidence $\theta=10^{\circ}$, actually the BIC relevant parameter, is fixed, so that the quasi-BIC regime is actually being explored here at a fixed point in parameter space; the resulting Q factor ($Q=2\omega_0/\Gamma$) is shown in Fig.\ref{fig:MD}c. The term $\sigma(\varepsilon)$ plays the role of the  background and basically depends on the broad MD$_{||}$ resonance. The asymmetry factor governing the quasi-BIC line shape is given by $q$ and  depends obviously on the MD$_{\perp}$ dressed polarizibility: since it is the crucial parameter in our analysis, it is explicitly shown in Fig.\ref{fig:MD}c. As expected, it can take on a variety of values both positive and negative. Importantly, from Eq.~(\ref{Eq:FanoCEMD}), it should vanish ($q=0$) when both MD resonances overlap. This is actually the case for $D/a\approx 0.75$, precisely the regime where full transparency ($R=0$, since $\varepsilon=0$ too) should emerge, as we will show in what follows. Recall that an \textit{asymmetry} factor $q=0$ in the Fano canonical formula actually indicates a \textit{symmetric} resonance in the form of a dip in a high background, as is the case of EIT.  

\section{\label{sec:FANOBIC}quasi-BIC-induced transparency band}

It is evident from Fig.~\ref{fig:MD}b,c that both MD resonances overlap for $D/a\sim 0.75$ leading to a narrow dip with an asymmetry factor $q\approx 0$; namely, a narrow EIT band. Therefore, let us calculate the reflection spectra from an infinite array for varying angle of incidence $\theta$ through our CEMD in such a case: recall that, for $D/a=0.5$, both MD resonances do not overlap and the MD$_{\perp}$ gives rise to a symmetry-protected (Brewster-like) BIC as revealed experimentally for $a=12$ mm in the microwave regime  \cite{Abujetas2019d}. The resulting reflection intensity and phase are shown in Fig.~\ref{fig:CEMD}a,b for $D/a\sim 0.75$. Near normal incidence, a broad reflection band is observed, with a narrow dip of negligible reflection splitting it, in turn tending to vanish at $\omega a/c\sim 1.6(2\pi)$. On the other hand, a similar behavior is observed on approaching grazing incidence at  $\omega a/c\sim 1.57(2\pi)$.
To shed light on the underlying physical mechanism, the contributions (intensity and phase) from MD$_{\perp}$ and MD$_{||}$ are shown separately in, respectively, Fig.~\ref{fig:CEMD}c,d and Fig.~\ref{fig:CEMD}e,f. 

Focusing on the normal incidence band of interest, we observe in Fig.~\ref{fig:CEMD}c that the MD$_{\perp}$ exhibits the characteristic behavior of a symmetry-protected BIC: namely, a narrowing resonant band with diverging Q-factor that disappears at normal incidence ($\Gamma$ point). The symmetry protection allowing BIC emergence was explained in terms of a Brewster-like mechanism \cite{Abujetas2019d}, forbidding  excitation/emission of a perpendicular dipole (either electric or magnetic) at normal incidence, otherwise the only available direction at the $\Gamma$ point imposed by the periodicity of the metasurface. However, the other  MD$_{||}$ resonance exhibits a broad band around normal incidence (see Fig.~\ref{fig:CEMD}e) due to the fact that coupling in (and scattering from) such dipolar resonance (with in-plane electric and magnetic fields) is fully permitted. The opposite occurs at grazing incidence: a BIC-like MD$_{||}$  resonance that interferes with the broad MD$_{\perp}$-resonant band. 
\begin{figure}
\includegraphics[width=\columnwidth]{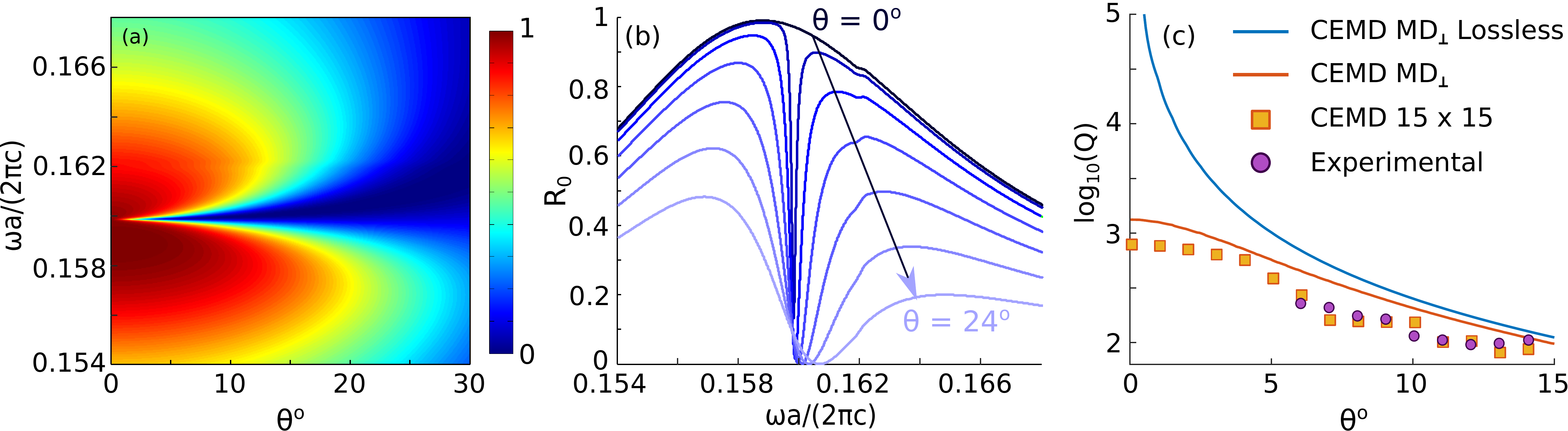}
\caption{(a) Contour map of the s-polarized reflectance $R(\omega,\theta$) intensity calculated through CEMD  from a square array ($D/a=0.75$) of HRI dielectric resonator disks (as in Fig. 2a but including only the contributions from the two MD resonances), zooming in the Fano-BIC region around $\omega a/c\sim 1.6(2\pi)$
and $\theta=0^{\circ}$;  (b) various spectra are shown explicitly for fixed $\theta$. 
(c) Q factors  obtained from the theoretical (reflectance)  spectral half width at half maximum in Eq.~(\protect{\ref{Eq:Fano}}), including separately that obtained from the MD$_{\perp}$ contribution only) of the BIT resonance, as a function of the angle of incidence $\theta$; expected Q-factor with losses ($Im(\epsilon)=0.05$)  accounted for, as in the experimental results shown below.
\label{fig:Fano}}
\end{figure}

The phase contour maps (see Fig.~\ref{fig:CEMD}d,f) confirm also such resonant behaviors through phase jumps. Incidentally, note also that interference of both MD resonances at intermediate angles of incidence leads to broad and dispersive bands of zero reflection (see Fig.~\ref{fig:CEMD}a), namely, Brewster-like bands with a peculiar phase dependence (see Fig.~\ref{fig:CEMD}b), similar to those described for dielectric cylinder metasurfaces in Ref.~\cite{Abujetas2018a}. On the other hand, it should be mentioned that there is also a weaker background at higher frequencies in Fig.~\ref{fig:CEMD}a stemming from the tail of higher-energy  electric-dipole resonances, responsible also for the Brewster-like band near grazing incidence at $\omega a/c\sim 1.73(2\pi)$.

Let us next analyze in more detail the interference between  MD resonance bands. This is shown by zooming in the angular/spectral BIC region  (see Fig.~\ref{fig:Fano}a), with reflectance spectra at various angles of incidence near the $\Gamma$ point included separately in Fig.~\ref{fig:Fano}b: it is evident from the  line shapes that the MD$_{\perp}$-BIC signature induces a narrow dip (quasi BIC-induced transparency, BIT) in the broad MD$_{||}$ band background, which has been achieved by tuning the metasurface periodicity to make overlap both MD resonances from isolated meta-atoms (disks). Figure~\ref{fig:Fano}c shows the corresponding angular dependence of the Q-factors, which diverge (in the absence of losses) upon approaching normal incidence, as expected, while saturating at nearly $Q\lesssim 10^3$ with absorption accounted for through $Im(\epsilon)=0.05$, corresponding to that of the mm disks used in the experimental results shown below.

\section{\label{sec:EXP}Experimental results}

\begin{figure}
\includegraphics[width=1\columnwidth]{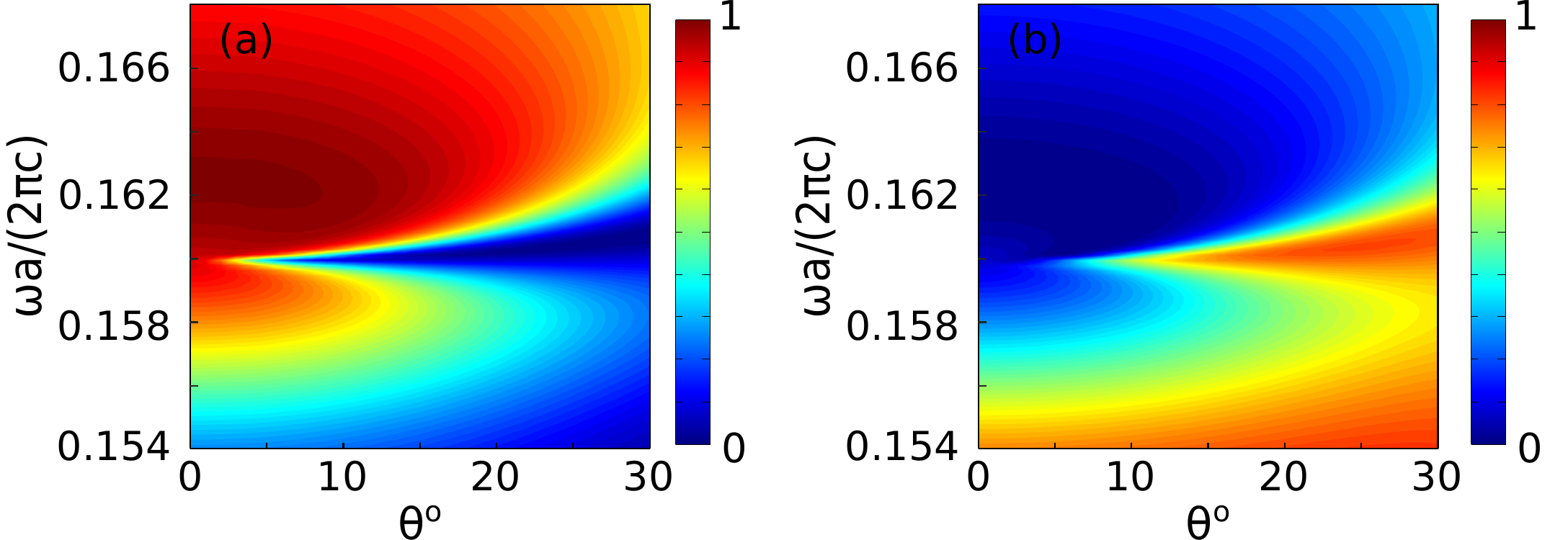}
\caption{Contour maps of the SCUFF numerical calculations of the  s-polarized (a) reflectance $R(\omega,\theta$) and (b)
and transmittance $T(\omega,\theta$) intensities  zooming in the BIT band 
for an infinite square array ($a=8$ mm) of dielectric resonator disks ($D=6$ mm and $L=4$ mm,  $\epsilon =78+ 0.05\mathrm i$) as a function of angle of incidence $\theta$  and normalized frequency $\omega a/(2\pi c)$. Normalized frequency $\omega a/(2\pi c)=0.16$ corresponds to $\nu=6$ GHz.
\label{fig:SCUFF}}
\end{figure}

As a proof of principle, let us  verify that our CEMD predictions hold for actual HRI disks. First,
numerical simulations have been carried out through SCUFF\cite{SCUFF1,SCUFF2}, showing the reflectance 
and the transmittance intensities
in s polarization in the angular/spectral region of the BIT band
for HRI disks ($\epsilon=78 + 0.05\mathrm i$ including weak absorption) with 6 mm diameter and 4 mm height, see  Fig.~\ref{fig:SCUFF}. The agreement  with the CEMD results (see Fig.~\ref{fig:CEMD}a) is not only qualitative, but also quantitatively good for all angles and frequencies
Only for grazing angles at higher frequencies (not shown) some discrepancies are observed near the above mentioned Brewster band, very likely due to higher multipolar contributions not accounted for in our CEMD method, which does underestimate reflection. 
More importantly, the BIT band is fully confirmed in transmission in a full numerical calculation accounting for real losses, as evidenced in Fig.~\ref{fig:SCUFF}b (recall that this was evident in the CEMD calculations without losses thanks to energy conservation being enforced). Note also that the near flat and narrow dispersion of the BIT band, stemming from the very nature of the "dark" mode responsible for it, namely, a BIC~\cite{Abujetas2019d}) , implies also a significant decrease of the group velocity that may lead to slow transmitted light~\cite{Schaafsma2016}.

\begin{figure}
\includegraphics[width=\columnwidth]{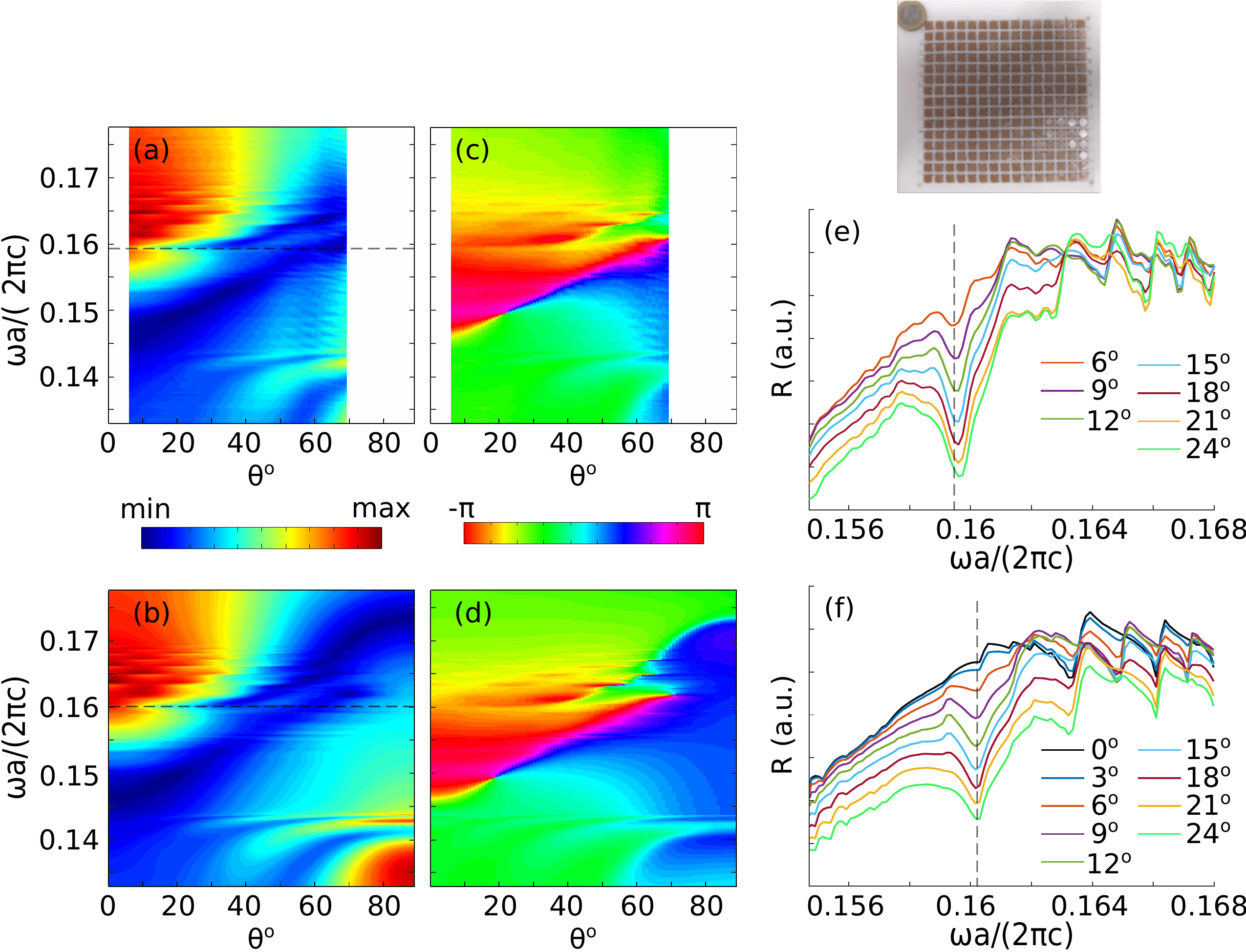}
\caption{Contour maps of the experimental measurements (a,c) and coupled dipole calculations (b,d) of the specular reflection intensity (a,b) and phase (c,d) as a function of angle of incidence $\theta$  and normalized frequency $\omega a/(2\pi c)$ for a finite square array (lattice period $a=8$ mm) of 15x15 disks ($D=6$ mm, $L=4$ mm, $\epsilon=78+ 0.05\mathrm i$). Spectra for fixed angles of incidence extracted from (a, respectively,b) are shown in (e, respectively, f). Dashed horizontal (a,b) and vertical (e,f) lines  identify the BIT band near the normalized frequency $\omega a/(2\pi c)\sim 0.16$, which corresponds to $\nu=6$ GHz. Top left panel: picture of the 15x15 disk sample.
\label{fig:EXP}}
\end{figure}

With both analytical (CEMD) and numerical evidence of the BIT band for infinitely extended all-dielectric metasurfaces, let us now show the experimental confirmation for a HRI-disk array in the microwave regime.  HRI disks are sold as a dielectric resonator of 6 mm  diameter and 4 mm height, with  permittivity given to be $\epsilon=78+ 0.05\mathrm i$. The finite array was made of an ensemble of 15x15 of those disks, positioned and spaced thanks to expanded polystyrene holders and spacers properly machined (as the permittivity of the expanded polystyrene is very close to one, its perturbation of the disk radiation is negligible);  see top right picture in \ref{fig:EXP}.
The measurements of the reflection  of the array were made in the anechoic chamber of the CCRM at the Institut Fresnel, the antennas being moved at a distance of about 2 m from the array. The antennas are linearly polarized, wide band ridged horn antennas, used here from 3.5 to 8.5 GHz and mechanically rotated along they axis to measure both s and p polarizations. Those measurements were made with or without the array, and calibrated with a metallic sphere as reference target. The source and receiving antennas were moved together and symmetrically to the normal of the array to cover an incident angular range of 6$^{\circ}$ to 70$^{\circ}$ in reflection (limitation due to the positioner capabilities); 
see Ref.~\cite{Abujetas2019d} and its supplementary information for more details. As those measurements are made with a network analyzer and as they are calibrated, the measured reflected and transmitted fields are determined in complex values, the amplitudes and phases thus both being directly comparable to theoretical calculations.

The main features  predicted by the analytical and numerical calculations above are  indeed observed in the experimental results in \ref{fig:EXP} (a,c,e), despite the fact that the fabricated, finite metasurface consists only of a limited number of disks). The expected quasi BIC-induced Fano resonance is  evident for low angles of incidence at the frequency within a broad band, 
revealing the expected BIT band at $\nu\sim$ 6 GHz, namely, $\omega a/(2\pi c)=0.16$, with decreasing Q-factor when approaching normal incidence, in agreement with the theoretical results. Other features mentioned above, such as the two zero-reflectance Brewster bands, are in turn measured experimentally. Moreover, even the expected phase jumps corresponding to those features in the reflectance intensity are also observed in all cases. Quantitative (not relevant)  discrepancies appear mostly at large (oblique) angles of incidence, at which the finite size of the sample should have a larger impact. To confirm that the slight discrepancies are indeed a finite size effect, we show also in Fig.~\ref{fig:EXP} (b,d,f) the reflectance (contour maps of intensity and phase, and spectra at fixed angles) calculated through a classical coupled-dipole theory  for a finite (15x15)  number of electric/magnetic dipoles; the agreement is qualitative and quantitatively excellent, especially for the BIT band.

This brings about the issue of the impact of  finiteness on the quality of the BIT band; recall that the  resonance width of a BIC should not tend to zero for a finite sample \cite{Taghizadeh2017,Abujetas2019d}. For the sake of comparison, Q-factors extracted from the experimental measurements and (finite) coupled-dipole theory  in Fig.~\ref{fig:EXP} are plotted in Fig.~\ref{fig:Fano}c, along with those predicted for an infinite array with and without absorption (the latter diverging, as expected for an ideal metasurface). The Q-factor for the finite sample indeed approaches $Q\sim 10^3$ near normal incidence, which is a very high value considering the size of the sample (only 15x15 unit cells). Interestingly, this value is remarkably close to that obtained for an infinite metasurface with losses accounted for (slightly above 10$^3$), yet another  evidence supporting the robustness of the  high-Q, quasi-BIC induced transparency band presented herein.

\section{\label{sec:Conc}Concluding remarks}
In summary, we have shown how to shift the spectral position of a canonical BIC in an all-dielectric metasurface, stemming from an isolated vertical MD resonance of the constituent meta-atoms (HRI disks), by tuning the periodicity of the array.  Indeed, we have analytically demonstrated through our coupled electric/magnetic dipole formulation that such  lattice-induced blueshift makes it overlap with the broad  (in-plane MD)  resonance, leading to a Fano resonance that becomes a narrow (quasi) BIC-induced transparency (BIT) band (the asymmetry factor vanishes) with a  diverging Q-factor typical of the canonical symmetry-protected BIC; the vanishing of the asymmetry factor ensuring BIT for overlapping MD resonances is in turn demonstrated analytically by rewriting the CEMD expression of the reflectance as a canonical Fano formula.  The emergence of a BIT band in the parameter space (angle of incidence) is evidenced through analytical and numerical calculations of the reflectance. Finally,  our experimental measurements in the microwave regime with a large array of  high-refractive-index disks have fully confirmed the theoretical predictions. Our results reveal a simple mechanism to engineer quasi BIC-induced transparency with arbitrarily large Q-factors that could be exploited throughout the electromagnetic spectrum with obvious applications in sensing, filtering, slow light, and non-linear optics.

\begin{acknowledgments}
J.A.S.G. and D.R.A. acknowledge partial financial support from the Spanish Ministerio de Ciencia e Innovaci\'on  (MICIU/AEI/FEDER, UE) through the grants MELODIA (PGC2018-095777-B-C21) and NANOTOPO (FIS2017-91413-EXP), and from the Ministerio de Educaci\'on, Cultura y Deporte through a PhD Fellowship (FPU15/03566). A.B. acknowledges the funding support from the Humboldt Research Fellowship from the Alexander von Humboldt Foundation. 
\end{acknowledgments}



\bibliography{library}

\end{document}